\documentclass[12pt]{elsart}
\usepackage{graphicx}

\newcommand{\gsim}{\stackrel{\scriptstyle >}{\phantom{}_{\sim}}}
\begin{document}
\begin{frontmatter}
\title{Thermal conductivity in dynamics of  first-order phase transitions}
\author[GSI]{V.V. Skokov }
and \author[GSI,MEPhI]{D.N. Voskresensky}
\address[GSI]{
GSI Helmholtzzentrum f\"ur Schwerionenforschung, D-64291
Darmstadt, Germany
}
\address[MEPhI]{Moscow Engineering Physical Institute,\\ Kashirskoe
  Avenue 31, RU-115409 Moscow, Russia}

\begin{abstract} Effects of thermal conductivity  on
the  dynamics of  first-order phase transitions are studied.
Important consequences of a difference of the isothermal and
adiabatic spinodal regions are discussed. We demonstrate that in
hydrodynamical calculations at non-zero thermal conductivity,
$\kappa \neq 0$,   onset of the spinodal instability occurs, when
the system trajectory crosses the
 isothermal spinodal line. Only for $\kappa =0$ it
occurs at a cross of the  adiabatic spinodal line. Therefore ideal
hydrodynamics is not suited for an appropriate description of
first order phase transitions.
\end{abstract}
\date{Nuclear Physics A on 26 Feb 2010}
\end{frontmatter}

\section{Introduction}

There are numerous phenomena, where first-order phase transitions
occur between phases of different  densities. In nuclear physics
various first-order phase transitions   may take place
 in  the Early Universe~\cite{Witten},
 in heavy ion collisions~\cite{Randrup,Shuryak:2008eq} and in neutron stars (e.g., pion,
kaon condensation and deconfinment or chiral phase transition~\cite{Glendenning,MSTV90}). At low collision energies the nuclear
gas-liquid (nGL)  first-order phase transition is
possible~\cite{Randrup,Ropke,BS,SVB,Siemens}. The experimental
status of the latter   is discussed in
Refs.~\cite{D'Agostino:1999kp,Schmidt:2000zs,D'Agostino:2005qj}.
At high collision energies  the hadron -- quark gluon plasma
(hQGP) first-order phase transition may occur (see e.g.
Ref.~\cite{Shuryak:2008eq}). Within the hydrodynamical approach
dynamical aspects of the nGL and hQGP  transitions have been
recently studied in Refs.~\cite{SV1,SV2,RandrupGQP}. An important
role of  effects of non-ideal hydrodynamics  has been emphasized
in  Ref.~\cite{SV1,SV2}. However in these works density
fluctuations were carefully analyzed, whereas thermal transport
effects were only roughly estimated.

Dynamical trajectories of the system in heavy-ion collisions can
be approximately determined by constant initial values of the
entropy per baryon (i.e.  entropy density per net baryon density).
This conclusion follows from  the analysis of  heavy ion
collisions at energies less than  several  GeV$/A$ performed in an
expanding fireball model (e.g., see Ref.~\cite{V93}), from ideal
hydrodynamical calculations  in a broad energy range (see e.g.
Refs.~\cite{Ivanov,Romatschke}) and also from transport
calculations (e.g., see Ref.~\cite{Arsene:2006vf}).

It is well known, see Refs.~\cite{RandrupGQP,Papp,FRS}, that, at
least in the mean field approximation, the isothermal spinodal
(ITS) line and adiabatic spinodal (AS) line are different. From
the exact thermodynamic relation one obtains
\begin{equation}
u_{\tilde{s}}^2  =u_T^2  + \frac{T}{nmc_V}   \left[ \left(\frac{\partial P}{\partial
T}\right)_n \right]^2 . \label{UC} %
\end{equation}%
Here $T$ is the temperature, $P$ is the pressure, $n$ is the
density of the conserving charge (here the baryon density), $m$ is
the baryon mass, $\tilde{s}\equiv s/n $ is the entropy per baryon,
$c_V$ is the specific heat density at the fixed volume $V$,
$u_{\tilde{s}}^2 =m^{-1}(\partial P/\partial n)_{\tilde{s}}$ and
 $u_T^2
=m^{-1}(\partial P/\partial n)_T$.  The value $u_{\tilde{s}}$ has
the meaning of the
 adiabatic sound velocity (at $\tilde{s} =$const). Namely, this quantity characterizes
 propagation of sound waves in ideal hydrodynamics. As we show
below, in non-ideal
 hydrodynamics at a finite value of the thermal conductivity, $\kappa$,
 the  propagation of sound waves is defined by
 the interplay between  $u_T$
 (at $T=$const) and $u_{\tilde{s}}$.
 Conditions $u_T =0$ and
$u_{\tilde{s}}=0$ define $T(n)$-curves: the ITS line (at fixed
temperature $T$) and AS line (at fixed $\tilde{s}$), respectively.
The maximum temperature points on these lines are the critical
temperature $T_{cr}$ in case of $T$=const and the adiabatic maximum
temperature $T_{P,max}$ in case of $\tilde{s}=$const. In the mean
field approximation, the specific heat density  $c_V$ has finite
non-negative values. Thus one has $u_T^2 \le 0$ on the AS line,
where $u_{\tilde{s}} =0$.

Calculations performed in  mean field models  show that $T_{cr}
\ge T_{P,max}$. In Refs.~\cite{BS,Siemens} and numerous  subsequent
works, crossing of the AS line by  effective trajectories
$\tilde{s}$=const  was considered as the starting point for
the multifragmentation following  an adiabatic expansion of the system.
In contrast,  in Ref.~\cite{SVB} and later Refs.~\cite{FRS,PR,LL} it was assumed
that
the multifragmentation may appear already at higher temperatures,
when the ITS curve is crossed by the $\tilde{s}$=const lines.
 The argumentation  of the
former group of works can be  traced back to the fact that in ideal
hydrodynamics instabilities indeed arise at the AS boundary (see
extensive discussions in Ref. \cite{RandrupGQP}). Isothermal
instabilities at first order phase transitions  were studied in
details in Refs. \cite{SV1,SV2}.

In mean field theories, difference between values of  $T_{cr}$ and
$T_{P,max}$ is usually rather large. For example, within the NJL model
one obtains  $T_{P,max} \sim T_{cr}/2 \simeq 45$~MeV  (see
Ref.~\cite{FRS})   for the chiral restoration
transition. In the model used in Ref.~\cite{RandrupGQP} $T_{P,max}
=3 T_{cr}/5 \simeq 100$~MeV for hQGP phase transition.
Note that simultaneously with the low value $T_{P,max} \sim
(50\div 100)$~MeV a sufficiently high value of $n_{P,max}$ should
be
reached. This  looks 
unlikely to be achieved in heavy-ion collisions.
Hence, if instability
occurs for $T<T_{P,max}$,
one could face
difficulties in exploring
the hQGP
 spinodal instabilities
in actual experiments with heavy ions.

It was argued in Ref. \cite{FRS}  that above  mentioned difference
of $T_{cr}$ and $T_{P,max}$ is probably an artifact of the mean
field approximation and both values of the critical temperature
should coincide provided thermal fluctuation effects are included.
Indeed, it is known (e.g., see ~Ref. \cite{helium}), that
inclusion of fluctuations may result in a weak divergence of the
specific heat. For example, for the three-dimensional  Ising
universality class $c_V \sim[(T-T_c)/T_c]^{-\alpha}$, where
$\alpha\simeq0.12$. It can be concluded from Eq.~(\ref{UC}), that
for divergent $c_V$ both temperatures  $T_{cr}^{\rm fl}$ and
$T_{P,max}^{\rm fl}$ should coincide. Although this observation is
fully true, it seems legitimate to rise some questions.
\begin{itemize}
\item
First, thermal fluctuation effects are strong enough only in a
critical region, where the mean field theory breaks down and
system properties are governed by non-trivial critical exponents.
For many systems the critical region is a rather narrow region
near the critical point, while   contrary examples are known as
well \cite{helium}. Although it is not known a priori how large is
the hQGP critical region, there is an indication in favor of the
hQGP critical region being small~\cite{Hatta:2002sj}. Only if the
system trajectory passes trough the critical region, one can hope
that fluctuation effects appearing during evolution of the system
may lead to $T_{cr}^{\rm fl} \simeq T_{P,max}^{\rm fl}$.
\item
Second, even if the system trajectory passes through the critical
region and $T_{cr}^{\rm fl}$ coincide with $T_{P,max}^{\rm fl}$,
the ITS and the AS lines may not collapse into one line  even
within the critical region because the fluctuation contribution to
the specific heat diverges only at the critical point and then
sharply decreases. Thus  we should understand crossing of which of
two lines (the ITS or AS line) results in the spinodal
instability.
\item
Third, the divergence of $c_V$ at the critical point  is due to
long-scale fluctuations, which need a long time to develop.
Therefore the time, which the system spends in a vicinity of the
critical point in the course of the heavy ion collision, might be
insufficient for the developing of such fluctuations, see
discussions in Refs.~\cite{SV1,SV2,Zeldovich,Berdnikov}.
\end{itemize}
Moreover,
if the initial collision energy is such that
the system trajectory passes outside  the critical region,
i.e. entirely in the mean field regime,
the thermal fluctuation effects do not contribute
at all. Then the question arises how one can treat the problem
purely within  the mean field approximation. This is actually
an important question, since most of equations of state used in the
description of heavy ion collisions disregard the
fluctuation effects.

In Refs. \cite{SV1,SV2} we focused on the study of dynamical
effects of the viscosity and the surface tension. The aim of the
given paper is to consider the influence  of  thermal conductivity
effects on the dynamics of first-order phase transitions and to
answer above risen questions.
 Below we find out  that in non-ideal
hydrodynamics (with non-zero thermal conductivity, $\kappa \neq
0$) instabilities show up at the crossing of the ITS line and only
for $\kappa =0$, e.g. in ideal hydrodynamics, they appear at
 the AS line.

\begin{figure}[t]
\centerline{
\includegraphics[height=6cm] {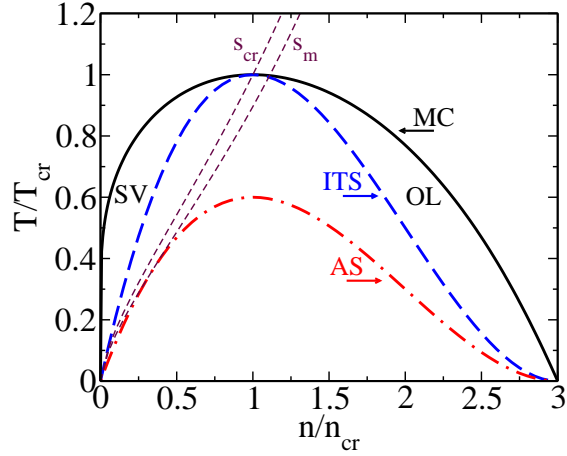}}
\caption{ \label{spinT} Phase diagram of the Van der Waals
equation of state, $T(n)$-plane. Bold solid,  dashed and
dash-dotted curves demonstrate   boundaries of the Maxwell
construction, the spinodal region at $T=$const and
$\tilde{s}=$const, respectively. The short dash lines show
adiabatic trajectories of the system evolution, see explanation in
the text.}
\end{figure}

\begin{figure}[t]
\centerline{
\includegraphics[height=6cm] {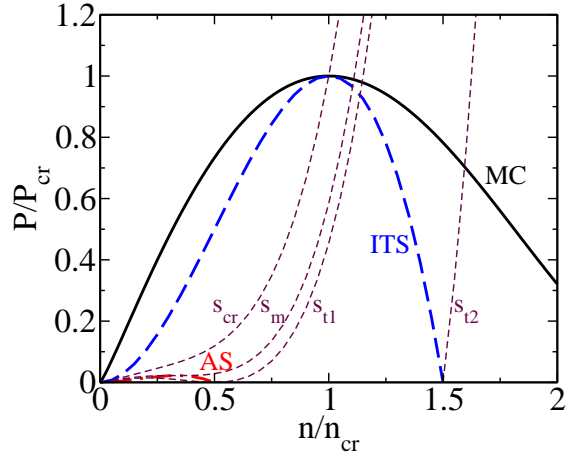}}
\caption{ \label{spin} Phase diagram for the Van der Waals
equation of state in $P(n)$ plane, see explanation in the text.
}
\end{figure}

\section{General setup}

To simplify further considerations we describe the  dynamics of
a first-order phase transition by means of the standard system
of equations of non-ideal non-relativistic
hydrodynamics\footnote{ A short survey of some  results
generalized for relativistic case is done in Appendix B.}:
the Navier-Stokes equation, the continuity equation, and the
general equation for the heat transport:
\begin{eqnarray}
\label{Navier} &&mn\left[ \partial_{t} {u}_i + (\mathbf{u}\nabla)
{u}_i \right]= -\nabla_i P \\ \label{contin}&& + \nabla_k \left[
\eta \left( \nabla_k u_i + \nabla_i u_k -\frac{2}{d} \delta_{ik}
\mbox{div} \mathbf{u}   \right)   + \zeta \delta_{ik} \mbox{div}
\mathbf{u} \right], \nonumber
\end{eqnarray}\begin{eqnarray}
&&\partial_{t}n +\mbox{div} (n \mathbf{u})=0,
\end{eqnarray}\begin{eqnarray}\label{therm}  && T\left[\frac{\partial
s}{\partial t} +\mbox{div}(s\mathbf{u} )\right]=\mbox{div}(\kappa
\nabla T)
\\&&
+\eta \left(\nabla_k u_i + \nabla_i u_k -\frac{2}{d} \delta_{ik}
\mbox{div} \mathbf{u} \right)^2 +\zeta (\mbox{div}
\mathbf{u})^2.\nonumber
\end{eqnarray}
Here
$\eta$ and $\zeta$ are the shear and bulk viscosities; $\mathbf{u}$ is the velocity of the fluid element;
$s$ is the entropy density; $\kappa$ is the thermal
conductivity;  $d$  is the dimensionality of space.
The short-scale fluctuations can be taken into account by adding  noise terms on the right
sides of Eqs.~(\ref{Navier}-\ref{therm}). In what follows we neglect these contributions.

Before we proceed further we aware the reader, that   Eqs.~(\ref{Navier}-\ref{therm})
 are derived in  the first gradient order approximation, see Ref.~\cite{Martin}. In
the kinetic regime the typical inverse relaxation time of the system $\tau_{\rm rel}^{-1}$ and the
typical inverse spatial scale $k_{\rm kin}$ characterizing the collective processes  fulfil the inequalities
$\tau_{\rm rel}^{-1} \ll  \epsilon_{\rm ch}$ and $k_{\rm kin}\ll p_{\rm
ch}$, see \cite{KB},
where  $\epsilon_{\rm ch}$ and $p_{\rm ch}$ characterize microscopic processes. For
 quasi-equilibrium nucleon system
with  $T\gsim \epsilon_{\rm F}$ ($\epsilon_{\rm
F}$ is the nucleon Fermi energy) one has $\epsilon_{\rm ch}\sim T$
and $p_{\rm ch}\sim p_T = \sqrt{2m T}$.
The hydrodynamic regime is reached if the typical time scale
and the typical length scale characterizing the collective motion
are larger than the corresponding kinetic values,
$\tau_{\rm hydro} \gg \tau_{\rm rel}$ and
$l_{\rm hydro}\gg 1/k_{\rm kin}$. In what follows we assume that these inequalities are fulfilled.

An approximate validity of the adiabatic expansion (i.e. conservation
of specific  entropy density, $\tilde{s}$, along the fluid flow) in
heavy ion collisions~\cite{Arsene:2006vf}
means that
(i) $\kappa$ is rather small, or the system is spatially
homogeneous and (ii) $\eta$ and $\zeta$ are small, or  the
velocity of the expansion of the system is very low (with the
guaranty for absence of shock waves). Note that in the general case
solutions of Ref. \cite{SV1}, which we will exploit below, are
obtained at a non-zero viscosity therefore we will not additionally
assume smallness of the viscosity coefficients, rather we exploit  smallness of the velocity $u$.

The best known example to illustrate principal features of a first
order phase transition in the mean field approximation is the Van
der Waals fluid. Expressions for thermodynamical quantities in
this particular case are presented in Appendix A. Trajectories for
the expansion of the uniform system, obtained at the assumption
$\tilde{s}\simeq$ const,  are shown in Fig. 1 on the plot
$T/T_{cr} =f(n/n_{cr})$. The upper convex  curve (MC, bold solid
line) is the boundary of the Maxwell construction, the next bold
dashed line (ITS) shows the boundary of the ITS region and the
lower bold dash-dotted curve (AS), the boundary of the AS region.
The short dashed lines crossing the regions under these three
curves are calculated at two  values of the entropy per baryon:
$\tilde{s}_{cr}$ and $\tilde{s}_{m}$. Different scenarios of the
dynamics for first-order phase transitions in expanding system at
$\tilde{s}\simeq$ const have been considered previously in Ref.
\cite{SVB}. The super-cooled vapor  (SV) and the overheated liquid
(OL) regions are between the MC (bold solid curve) and the ITS
(bold dash) curves, on the left and on the  right respectively.
For $\tilde{s}_{cr}>\tilde{s}>\tilde{s}_{\rm MC2}$, where
$\tilde{s}_{cr}$ corresponds to the critical point and the line
with $\tilde{s}_{\rm MC2}$ passes through the point $n/n_{cr}=3$
at $T=0$, the system traverses the OL state (region OL in Fig. 1)
and then the ITS region (below ITS line) and the AS region (below
AS line). For $\tilde{s}>\tilde{s}_{cr}$ the system trajectory
passes through the SV state (the region SV in Fig. 1) and then the
ITS region.

In Fig. 2 we show the same regions as in Fig.1, but in the plane
$P/P_{cr} - n/n_{cr}$ for $P>0$. The line $\tilde{s}_m$
corresponds to the maximum pressure at the AS curve. The critical
point is determined by $(dP/dn)_{T}=0$, $(d^2 P/dn^2)_{T}=0$. With
the help of Eq. (\ref{P}) of Appendix A from these conditions we
find $n_{cr} =1/(3b)$, $T_{cr} =8a/(27b)$. The maximal pressure
point at AS line is defined by $(dP/dn)_{\tilde{s}}=0$,  $(d^2
P/dn^2)_{\tilde{s}}=0$. Thus using Eq. (\ref{Ps}) we obtain:
$n_{P,max} =1/(9b)=n_{cr}/3$, $T_{P,max} =128a/(1215b)=16 T_{cr}
/45$, see Fig. 2. The maximal temperature point on the AS curve
corresponds to $T_{max}=3/5 T_{cr}$, $n_{max}=n_{cr}$, see Fig. 1.
Another relevant point on each line $\tilde{s}={\rm const}$ is the
turning point, where $E(n, \tilde{s}={\rm const})$ has a minimum as
a function of the density, $n$.  In this point the pressure $P(n,
\tilde{s}={\rm const})$ equals zero and the fireball expansion is
slowing down.  The lines labeled $\tilde{s}_{t1}$ and
$\tilde{s}_{t2}$ cross respectively the AS and
 ITS lines  at $P=0$. Note, that with realistic equation of state,  $P=0$ might be
reached at the nGL first-order phase transition, see
Ref.~\cite{Papp}, but not at the hQGP transition.

We stress that we use the Van der Waals equation of state only for the illustrative
purpose.  Our main conclusions do not depend on the used equation of state provided
there is  a first order phase
transition.

We may expand all quantities near  an arbitrary reference point
$(n_{\rm r}, T_{\rm r})$, or  $(n_{\rm r}, \tilde{s}_{\rm r})$.
Analytical solutions can be found, provided the system is in the
vicinity of the critical point $(n_{cr} , T_{cr})$, or near the
maximum point $(n_{P,max},\tilde{s}_{P,max})$. Therefore
considering the problem in $(n,T)$ variables it is convenient to
take $(n_{cr} , T_{cr})$ as the reference point and considering
the problem in $(n,\tilde{s})$ variables, to take $(n_{P,max} ,
\tilde{s}_{P,max})$ as the reference point. The following remark
is in order. As we have mentioned, in this paper we exploit the
mean field approximation. Thus using $(n_{cr} , T_{cr})$ as the
reference point we actually assume that the fluctuation  region
($n_{cr}\mp \delta n^{\rm fl}$, $T_{cr}\mp \delta T^{\rm fl}$) is
very narrow and we are able to chose the reference point outside
this region (but nevertheless very close to the critical point).
Also, mentioned problem is avoided provided one considers the
system at the time scale, being shorter than that necessary for a
development of long-scale critical fluctuations.

We construct the generating functional in ($\delta n
=n-n_{cr}$, $\delta T=T-T_{cr}$) or in ($\delta n=n-n_{P,max}$,
$\delta \tilde{s}=\tilde{s}-\tilde{s}_{P,max}$)
variables, the Landau free energy,
such that $\delta (\delta F_L)/\delta(\delta n) = P - P_{f}+P_{\rm
MC} $:
\begin{eqnarray}\label{fren}
&&\delta F_L = \int \frac{d^3 x}{n_{cr}}\left\{
\frac{cm[\nabla (\delta n)] ^2}{2}+\frac{\lambda m^3 (\delta
n)^4}{4}-\frac{\lambda v^2 m(\delta n) ^2}{2}-\epsilon \delta n
\right\},
\end{eqnarray}
where $\epsilon = P_f-P_{\rm MC}$ is expressed through the pressure on
the Maxwell construction. The maximum of the quantity $\epsilon$ is
${\epsilon}^{m}= 4\lambda v^3 /(3\sqrt{3})$, therefore we may
count ${\epsilon}$ from this value supposing further that
${\epsilon}=\pm \gamma_{\epsilon}{\epsilon}^{m}$ with
$0<\gamma_{\epsilon}<1$.
The first term in Eq. (\ref{fren}) is due to the surface tension.
It produces a surface pressure $P_{surf}$ (see Eq. (\ref{v-t})
below).

In  case of the Van der Waals form of the equation of state, see
Appendix A, expanding the pressure of the uniform matter in $\delta
n=n-n_{cr}$, $\delta T=T-T_{cr}$, for $|\delta n| /n_{cr} \ll 1$,
$|\delta {T}|/T_{cr} \ll 1$, i.e. in the vicinity of the critical
point $(n_{cr} ,T_{cr})$, we obtain
\begin{equation}\label{modif}
\delta P[n,T]=f_0 \left[\frac{9\delta{{ T}}\delta n
}{4}+\frac{9T_{cr} (\delta n)^3}{16n_{cr}^2}  + \frac{3}{2} n_{cr}
\delta T +...\right].
\end{equation}
The last explicitly presented term in Eq.~(\ref{modif}) is unimportant
if $T(\vec{r})\simeq$ const, since the addition of any constant
does not change equation of motion~(\ref{Navier}).

Comparing Eq.~(\ref{modif}) with the value, which follows from Eq.
(\ref{fren}),
we find relations between coefficients:
\begin{eqnarray}\label{parame}
&&v^2 (T) =- 4 \delta{{T}} n_{cr}^2 m^2/T_{cr} ,\nonumber\\
&&\lambda_{cr} =\frac{9f_0 }{16}\frac{T_{cr}}{n_{cr}^2 m^3}.
\end{eqnarray}

Expanding the pressure with respect to  $\delta n =n-n_{P,max}$,
$\delta\tilde{s}=\tilde{s}-\tilde{s}_{P,max}$ in the
vicinity of the maximum of $P(n)$ at fixed ${\tilde{s}}$ we
obtain
\begin{eqnarray}\label{modifS} &&\delta
P[n,\tilde{s}]=\frac{15}{512}T_{P,max}
\left[ \delta \tilde{s} \delta n +\frac{f_0 (\delta
n)^3 }{2
n_{P,max}^2}+\frac{114}{15}n_{P,max}\delta\tilde{s}+...\right].
\end{eqnarray}
Comparing this expression with the value, which follows from
 Eq.
(\ref{fren}),
we find relations
\begin{eqnarray}\label{parameS}
&&v^2 (\tilde{s}) =- 2 \delta{\tilde{s}} n_{P,max}^2
m^2/f_0 ,\nonumber\\ &&\lambda_{P,max} =\frac{15f_0
}{1024}\frac{T_{P,max}}{n_{P,max}^2 m^3}.
\end{eqnarray}

Further, following Refs. \cite{SV1,SV2}, we use smallness of the
velocity $\mathbf{u}(t,\mathbf{r})$ of the growth/damping of the
density and temperature fluctuations. We linearize all terms in
hydrodynamical equations in the velocity $\mathbf{u}\,$, the
entropy density $\delta s=s-s_{\rm r}$ and the temperature $\delta
T=T-T_{\rm r}$ variables. Here index  ``r'' stays for  an
arbitrary reference point.

Excluding a dependence on the velocity, $\mathbf{u}$, from
Eqs.~(\ref{Navier}), (\ref{contin})  we obtain~\cite{SV1,SV2,V93a}
\begin{equation}\label{v-t}
m\partial^2_t \delta n
=\Delta \left[P+P_{surf} -P_{f}
+n_{\rm r}^{-1}\left(\tilde{d}\eta_{\rm r} + \zeta_{ \rm r}
\right)
\partial_t \delta n
\right] ,
\end{equation}
where  $P_{surf}$ is the surface pressure;
$\tilde{d}=2(d-1)/d$, $\tilde{d}=4/3$ for spherical
geometry, $\delta n=n-n_{\rm r}$. We count $P$ from its value in
the final equilibrium state (reached at $t\rightarrow \infty$).
Namely this difference, $  P - P_{f}$, has the meaning of the
thermodynamical force driving the system to the final equilibrium
state.

Eq.~(\ref{v-t}) should be supplemented by the equation for the heat
transport (\ref{therm}), which after linearization owing to Eq.~(\ref{contin})  simplifies as
\begin{equation}\label{v-tS}
T_{\rm r}\left[
\partial_t \delta s
-s_{\rm r}(n_{\rm r})^{-1}
\partial_t \delta n
\right]=\kappa_{\rm r}\Delta \delta T.
\end{equation}
The variation of the temperature is related to the variation of
the entropy density $s[n,T]$ by
\begin{equation}\label{T-s}
\delta T \simeq T_{\rm r} (c_{V ,\rm r})^{-1}\left(\delta s
-({\partial s}/{\partial n})_{T,\rm r}\delta n\right) ,
\end{equation}
where, as in Eq.~(\ref{UC}), $c_{V ,\rm r}$ is the specific heat
density.

\section{Instabilities in spinodal region}\label{Instabilities}

\subsection{Effect of finite thermal conductivity} Since in this
section the ``r''-reference point can be taken arbitrary, we
suppress index ``r''. To find solutions of hydrodynamical
equations we present, cf. \cite{SV1,SV2,RandrupGQP},
\begin{eqnarray}\label{delns}
\delta n =\delta n_0 \mbox{exp}[\gamma t +i \mathbf{p}\mathbf{r}], \nonumber \\
\delta s=
\delta s_0 \mbox{exp}[\gamma t +i \mathbf{p}\mathbf{r}], \\
T =T_{>}+\delta T_0 \nonumber
\mbox{exp}[\gamma t +i \mathbf{p}\mathbf{r}].
\end{eqnarray}
where $T_{>}$ is the temperature of the uniform matter.

 For finite
thermal conductivity, $\kappa \neq 0$, from  Eqs. (\ref{v-tS}),
(\ref{T-s}) we obtain
\begin{eqnarray}\label{123}
&&\delta s_0 =\delta n_0\frac{ s  }{ n \left[1+\kappa p^2/(c_{V
}\gamma)\right]} \left[1+\frac{ n \kappa
p^2(\partial s/\partial n)_{T} }{\gamma s c_{V }}\right].
\end{eqnarray}
 The increment,
$\gamma (p)$, is determined by Eq. (\ref{v-t}), where we put
 \begin{equation}\label{varP}
 \delta P= \left( \frac{\partial P}{\partial n} \right)_{T}\delta n +\left( \frac{\partial P}{\partial T} \right)_{n}\delta
 T .
 \end{equation}
 Thus we obtain
 \begin{eqnarray}\label{g2} &&\gamma^2 =
-p^2 \left[u_T^2 +\frac{(\tilde{d}\eta +\zeta)\gamma
}{mn}+cp^2 +\frac{u_{\tilde{s}}^2 -u_T^2 }{1+\kappa p^2
/(c_{V}\gamma)}\right]. \end{eqnarray}
  This equation differs from that has been derived
in Ref.~\cite{PR} by presence of extra surface tension term ($\propto
c$).

Eq. (\ref{g2}) has three solutions. Expanding the solutions for small
momenta (long-wave limit) we find
\begin{eqnarray}
\gamma_{1,2} &=& \pm i u_{\tilde{s}} p + \left[
\frac{\kappa}{c_{V }} \left(
\frac{u_T^2}{u_{\tilde{s}}^2}-1\right)
 - \frac{\tilde{d}\eta +\zeta }{mn}
\right]\frac{p^2}{2},\label{solG1}   \\
\gamma_3 &=& - \frac{\kappa
u_T^2 p^2}{u_{\tilde{s}}^2c_{V }} \left[ 1 -
\frac{u_T^2-u_{\tilde{s}}^2}{u_{\tilde{s}}^2u_T^2}\left( c
+ \frac{\kappa u_T^4 }{u_{\tilde{s}}^2 c_{V }^2}
   - \frac{(\tilde{d}\eta +\zeta) \kappa u_T^2}{mn
c_{V } u_{\tilde{s}}^2 } \right) p^2 \right]. \label{solG}
\end{eqnarray}
The solutions $\gamma_{1,2}$ correspond to the sound mode in the
long wavelength  limit. The third solution, $\gamma_3$, is responsible
for the thermal transport mode. Below the ITS line (and above the
AS line) $u_T^2 <0$, $u_{\tilde{s}}^2
>0$, solutions $\gamma_{1,2}$ correspond to an oscillation and damping, whereas  $\gamma_3$ describes an unstable
growing solution. Below the AS line, since $u_{\tilde{s}}^2<0$ and
$u_{T}^2<0$, the modes exchange their roles: the sound modes
become unstable, while the thermal mode is damped.

We solved the system of non-relativistic hydrodynamical equations.
Solutions generalizing Eqs.~(\ref{solG1}) and (\ref{solG}) to the
relativistic case are presented in Appendix B.

 Further  we will
consider two limiting cases of a large and small thermal
conductivity, $\kappa$.

\subsection{Limit of a large thermal conductivity}

 For $\kappa
p^2 /(c_{V }|\gamma|)\gg 1$ from Eq.~(\ref{g2}) we arrive at
 \begin{equation}\label{g} \gamma^2 =
-p^2 \left[u_T^2 +\frac{(\tilde{d}\eta +\zeta )\gamma
}{mn}+cp^2\right],
\end{equation}
that we have derived in Refs. \cite{SV1,SV2}.
Note that dependence on $\kappa$ disappeared in this
limiting case. The solutions are
 \begin{equation}\label{gsol}
 \gamma =-p^2\sqrt{\frac{c}{4\beta}}\pm
 \sqrt{cp^4\left(\frac{1}{4\beta}-1\right)-p^2u_T^2}.
\end{equation}
 Here $\beta =c(mn)^2/ (\tilde{d}\eta
+\zeta)^2$ is the dimensionless parameter.
For $c|u_T|/\kappa \ll p\ll
|u_T|/\sqrt{c|1+1/\beta|}$ we obtain $\gamma^2 =-u_T^2 p^2$, and
$u_T^2 =-\lambda_{cr} v^2 (T
)=9\delta T f_0 /(4m)$ for the particular case of the Van der
Waals fluid in the vicinity of the critical point $(n_{cr},
T_{cr})$ (taken as the reference point, cf.  Eq. (\ref{modif})).

An instability arises  within the ITS region, where $u_T^2 <0$. The
most rapidly growing mode corresponds  to
 $\gamma \simeq
\gamma_{m}$, and $p\simeq p_{m}$:
\begin{eqnarray}\label{gmax}
&&\gamma_{m}
=\frac{(-u_T^2)mn_{cr}
 }{(2\sqrt{\beta}+1)(\tilde{d}\eta
+\zeta)},\\ &&p_{m}^2
= \frac{(-u_T^2)\sqrt{\beta}}{(2\sqrt{\beta}+1)c}. \nonumber
\end{eqnarray}
  Using these
expressions we rewrite the condition $\kappa p^2 /(c_{V}\gamma)\gg
1$ as $$ \kappa\gg c_{V } \sqrt{c}.$$
Then also we have
\begin{eqnarray}
&&1+\frac{ n \kappa p^2_{m}(\partial s/\partial n)_{T}}{\gamma_m s
c_{V }}=1+\frac{\kappa n(\partial s/\partial
n)_{T}}{s\sqrt{c}c_{V}},\nonumber\\ &&1+\kappa
p^2_{m}/(c_{V}\gamma_m)=1+ \frac{\kappa}{c_{V }\sqrt{c}},
\end{eqnarray} and following (\ref{123}):
\begin{equation}\label{T0}
\delta T_0 =\delta n_0 \frac{T s [1-n(\partial s/\partial
n)_{T}/s]}{c_{V } n \left[1+{\kappa} /(\sqrt{c}c_{V } )\right]}.
\end{equation}
 We see that an assumption on the spatial uniformity of the
system fails right after  the ITS region is reached. An aerosol
(mist) of bubbles and droplets is formed for a typical time
$t_{aer}\sim 1/\gamma_{m}$.  For the Van der Waals equation of
state using Eq. (\ref{S}) of Appendix A   we find that $\delta T_0
/\delta n_0
>0$, i.e. the temperature is larger in denser regions. However
the amplitude of the temperature modulation is rather small for $
\kappa\gg c_{V } \sqrt{c}.$

In the limiting case,  one can neglect at all small deviations of $\delta T_0$
 and put $\delta T_0 =0$ in Eq. (\ref{T0}).
 Formally this limit corresponds to setting  $\kappa \rightarrow
\infty$ in Eq.~(\ref{T0}). In order to avoid misunderstanding we stress that we continue to
consider only slow motions staying in the framework of the validity of the
 non-relativistic hydrodynamics~\cite{Martin}.

As before,
we assume that at a slow expansion with $\tilde{s}(t)\simeq$ const
the initially uniform system
 reaches   ITS  region
(see corresponding curves in Fig. 2).
Solution of Eq. (\ref{therm}) for $\kappa \rightarrow \infty$
corresponds to
 $T (\mathbf{r}) ={\rm const}$,
  and to the approximate conservation of
the entropy  (provided the velocity $u$ or the viscosities are
small), as it follows from the l.h.s of Eq. (\ref{therm}). We may
also see it from Eq. (\ref{v-tS}). For that let us divide this
equation on $\kappa$. Then we can put zero the l.h.s. of thus
obtained equation   (since $\kappa \rightarrow \infty$). Any
solution $T (\mathbf{r}) ={\rm const}$ fulfils this equation. Using
Eq.~(\ref{delns}) we arrive at  $\delta T =0$ (not dependent on
$\mathbf{r}$ and $t$).
 The latter approximation was used in
 Refs. \cite{SV1,SV2}, where we solved Eq. (\ref{v-t}) for $T=$const and derived  Eq. (\ref{g}) and  Eq. (\ref{gsol}).
On the other hand, from Eq.
(\ref{T-s}) at $\delta T=0$  we find
$\delta s_0 =({\partial s}/{\partial n})|_{T}\delta n_0$,
that follows from Eq.~(\ref{123}) as well.
 Using that $\delta T=0$ in Eq.~(\ref{varP}) we again arrive at Eq.~(\ref{g}) and  Eq.~(\ref{gsol}).

\subsection{Limit of small thermal conductivity}
For the sake of simplicity,
here we consider the case, when both shear and bulk
viscosities are zero, but thermal conductivity is small but
non-zero. Dropping $\gamma^2$ term in the l.h.s. of Eq. (\ref{g2})
(we justify this approximation below)  we immediately find the
solution
 \begin{equation}
 \gamma_3 =-\frac{\kappa u_T^2 p^2}{u_{\tilde{s}}^2
 c_V}\frac{1+cp^2/u_T^2}{1+cp^2/u_{\tilde{s}}^2} .
\end{equation}
Expanding this value in small $p$ we see that it, indeed,
corresponds to the solution $\gamma_3$  given by  Eq.
(\ref{solG}).

 The most rapidly growing mode corresponds to
 \begin{eqnarray}
p_m^2 =\frac{u_{\tilde{s}}^2}{c}\left(-1+\sqrt{1-\frac{u_T^2}
{u_{\tilde{s}}^2} }\right)>0.
\end{eqnarray}
Slightly below  the ITS line (for $-u_T^2\ll 1$),
 \begin{eqnarray}\label{gam3m}
p_m^2 \simeq -u_T^2 /(2c), \quad  \gamma_{3m} =\gamma_3
(p_m)\simeq \frac{\kappa u_T^4}{4cc_V u_{\tilde{s}}^2}.
 \end{eqnarray}
Now we are able to check the validity of our solution in this
case. Comparing different terms in Eq.~(\ref{g2}) we see that it  is
legitimate to drop the term $\gamma^2$ in the l.h.s. of equation,
provided
 \begin{eqnarray}
 \kappa \ll c_V \sqrt{c}\frac{u_{\tilde{s}}^2}{|u_T^2|}.
 \end{eqnarray}
Comparing Eq.~(\ref{gam3m}) with Eq.~(\ref{gmax}) for $\beta \rightarrow
\infty$ (zero viscosity) we see that at $c_V \sqrt{c}\ll \kappa
\ll c_V \sqrt{c}{u_{\tilde{s}}^2}/{|u_T^2|}$, when both
solutions $\gamma_m$ and $\gamma_{3m}$ are valid,  the following inequality is satisfied $\gamma_m \gg
\gamma_{3m}$.

 Thus we found that in all considered above cases the instability occurs
at the crossing of the ITS line. However, for $\kappa \gg c_V \sqrt{c}$ the
most rapidly growing mode corresponds to $\gamma_m$ and in the
opposite limit  $\kappa \ll c_V \sqrt{c}$, to $\gamma_{3m}$.

\subsection{Limit of $\kappa =0$} In this specific limit there
are significant differences compared to the case of any non-zero
$\kappa $. Indeed, for $\kappa =0$ there is no solution with
$\gamma_3 \neq 0$.
Below the ITS line and above the AS line
the thermal mode, that drives the system towards equilibrium
for non-zero thermal conductivity, does not exist for  $\kappa =0$.
Thus the evolution in the spinodal region is entirely governed by
approximately adiabatic sound excitations.

From
Eq.~(\ref{v-tS}) we find
\begin{equation}\label{s0}
 \delta s_0 =\delta n_0 s /n ,
 \end{equation}
and from Eqs.~(\ref{T-s}) and (\ref{s0})
 \begin{equation}\label{To} \delta
T_0 =\delta n_0 \frac{T s [1-n(\partial s/\partial
n)_{T}/s]}{c_{V} n }. \end{equation}
 Thereby  the temperature is modulated, similar to the
entropy density. For the Van der Waals equation of state using
Eq.~(\ref{S}) of Appendix A   we find that $\delta T_0 /\delta n_0
>0$. The amplitude of the
 temperature modulation  is here larger than in case of large $
\kappa$, see Eq. (\ref{T0}).

The increment $\gamma$ is given by Eq. (\ref{g})  with
$u_{\tilde{s}}$ replaced by $u_T$:
 \begin{equation}\label{g1} \gamma^2 =
-p^2 \left[u_{\tilde{s}}^2 +\frac{(\tilde{d}\eta
+\zeta)\gamma }{mn}+cp^2 \right].
\end{equation}
Thus, contrary to the case $\kappa \neq 0$, instability appears,
when the system trajectory crosses the AS line rather than the ITS
line. The value $u_{\tilde{s}}^2 =-\lambda_{P,max} v^2
(\tilde{s}
)=15\delta \tilde{s} T_{P,max} /(512 m)$ for the particular
case of the Van der Waals fluid in the vicinity of the point
$(n_{P,max}, T_{P,max})$ (taken as the reference point, cf. Eq.
(\ref{modifS})). This result holds also in case of the  ideal
hydrodynamics, where in addition to  $\kappa =0$, the viscosity coefficients ($\eta$, $\zeta$) are  zero.

Note that for any small but finite value $\kappa$ the solution
$\gamma_3$ results in instability already for  $u_T^2<0$ (i.e.
below the
ITS line rather than below the AS line).  Since in reality
always
$\kappa\neq 0$, we conclude that the instability condition
$u_{\tilde{s}}^2<0$, as one usually derives it
 within the framework of ideal hydrodynamics, should be replaced by
the condition  $u_T^2<0$.

\section{ Phase transition from metastable state}

During its evolution a dynamical system crosses  a metastable
state region (binodal) before entering to the region of the instability (spinodal), see Figs. 1 and 2.
If a system evolves rather slowly  a phase transition occurs already in a metastable region.
Therefore let us 
consider  the system  in the metastable
state.
Small  seeds of the stable phase, being produced in
fluctuations, converge, while  seeds of  overcritical sizes start to
nucleate, see Ref.~\cite{LLkin}. In nuclear physics these processes have been considered
in many works, e.g.  see Ref.~\cite{Randrup} and references therein.

In Refs.~\cite{SV1,SV2}  we  described  the seeds dynamics using
equations of non-ideal hydrodynamics. Here we continue this
description.
An analytical
description of metastable states  is possible to perform in the
limiting cases of infinitely large thermal conductivity (when one can put
$T=$const, see Refs.~\cite{SV1,SV2}) and of zero thermal conductivity.
Therefore, below we study the limiting cases at first and  then we perform  estimates
for an  arbitrary value of thermal conductivity,  $\kappa$.

\subsection{Limit of $\kappa\rightarrow \infty$}

 Let us consider the situation, when   at very slow expansion with
$\tilde{s}(t)\simeq$ const  spatially uniform spherical system of
a large radius $R_V$ achieves either the OL- state or the SV-
state (see corresponding curves in Fig. 1). During subsequent slow
expansion, after a while a bubble of an overcritical  size $R>R_{cr}$  of
stable gas phase, or respectively a droplet of  liquid  phase may
appear in a fluctuation and begin to grow. In case $|n -n_{cr}
|/n_{cr} \ll 1$, i.e. in the vicinity of the critical point
$(n_{cr},T_{cr})$, corresponding solutions (\ref{v-t}) describing
dynamics of the density in the fluctuation  were searched in Ref.
\cite{SV1} in the form :
 \begin{equation}\label{delr} \delta n (t,r)\simeq
\frac{v(T)}{m}\left[\pm\mbox{th} \frac{r-R_{n}
(t)}{l}+\frac{{\epsilon}}{2\lambda_{cr} v^3(T)}\right]+(\delta
n)_{cor},
\end{equation}
where $l=\sqrt{2c/(\lambda_{cr} v^2(T))}$, the upper sign
corresponds to the growth  of the bubble and the lower one, to the
growth of the droplet,
and the solution is valid for $|\epsilon/(\lambda_{cr} v^3(T))|\ll
1$, i.e. for $\gamma_{\epsilon} =$ const $ \ll 1$ and for the initial bubble/droplet size $R_n (0)
>R_{cr}(T)$.
 In Refs. \cite{SV1,SV2} we neglected a small correction $(\delta
n)_{cor}$ in Eq.~(\ref{delr}). With this small correction
one is able to recover  exact baryon number conservation, see
below. One can easily see that
Eq.~(\ref{delr}) fulfils Eq.  (\ref{v-t}) provided the temperature, ${T} $, does not depend on
time and space.
As we argued above, this result follows from Eq.~(\ref{v-tS}),
if one formally sets  $\kappa \rightarrow \infty$.

  In case of expanding
system we also assume that typical time for the formation and
growth of a fluctuation  of our interest ($t_{form}+t_{\rho}$) is
much smaller than the typical expansion time $t_{exp}$.

Substituting Eq.~(\ref{delr}) to  Eq.~(\ref{v-t}) and considering $r$ in
the vicinity of the bubble/droplet boundary we obtain equation
describing growth  of the bubble/droplet size \cite{SV1,SV2}:
\begin{equation}\label{dim}
\frac{\beta t_0^2}{2}
\frac{d^2R_{n}}{dt^2}=\frac{3\epsilon}{2\lambda_{cr} v^3 (T)
}-\frac{2l}{R_{n}}-\frac{t_0}{l}\frac{d R_{n}}{dt},
\end{equation}
where  $t_{0} =2(\tilde{d}\eta_{cr} +\zeta_{cr})/[\lambda_{cr}
v^2 (T) mn_{cr}]$ is the  time scale. It follows from the  solution of this
equation
 that $R_{cr}=4l\lambda_{cr}v^3(T)/(3\epsilon)$.
First the bubble/droplet size $R_{n}(t)>R_{cr}$ grows with an
acceleration and then it reaches a steady grow regime with a   constant velocity
$u_{as}=\frac{3\epsilon l}{\lambda_{cr} v^3 (T) t_0}\propto
\gamma_{\epsilon}|T_{cr} -T|^{1/2}$.

 In the
bubble/droplet interior $\delta n \simeq \mp v(T)/m$, and this
region grows with  time. Thus  $(\delta n)_{cor} \simeq
v(T)R_{n}^3(t)/ (mR_V^3)$ produces  compensation of the baryon
number. Note that this correction is very small for $R_{n}(t)\ll
R_V$ and does not affect solution at $R\sim R_{n}(t)$.

Substituting Eq.~(\ref{delr}) to Eq.~(\ref{T-s})
for
$T=$ const we obtain
 \begin{eqnarray} \label{deltasinf}
&&\delta s =
\left(\frac{\partial s}{\partial n}\right)_{T}\left\{\frac{v(T)}{m}\left[\pm\mbox{th} \frac{r-R_{n}
(t)}{l}+\frac{{\epsilon}}{2\lambda_{cr} v^3 (T)}\right]+(\delta
n)_{cor}\right\}.
 \end{eqnarray}
Thus the temperature stays constant in the bubble/droplet interior and
exterior and the entropy  and the density  differ in the
interior and exterior regions. In spite of that,  approximation of
a quasi-adiabatic expansion of the system as a whole might be used
even, when the system reaches metastable region, provided the gas
of bubbles/droplets is rare, or the value $v(T)$ is small.

\subsection{Effect of finite thermal conductivity}

Let us determine, when in the case $\kappa \neq 0$ we are still able
to exploit solutions found above for $\kappa \rightarrow\infty$.
Estimating from Eq. (\ref{T-s}) $\delta T\sim \delta s T_{cr}
/c_{V ,cr}$ (for $|\delta s|\sim |(\partial s/\partial
n)_{T}\delta n|$) with $\delta n$ from Eq. (\ref{delr}) and
setting this estimate in Eq. (\ref{v-tS})
we find the typical time for the heat transport $t_T \sim R^2 c_{V
,cr}/\kappa_{cr} $.   For $R(t)<R_{\rm fog}=\kappa_{cr}/(u_{as}
c_{V ,cr})$ one has $t_n > t_T$ and the droplet/bubble dynamics is
determined by Eq. (\ref{dim})  for the change of the density, cf.
\cite{SV1,SV2}. At $\kappa \rightarrow \infty$  for all times the
droplet/bubble dynamics is determined by Eq. (\ref{dim}).  At
$\kappa\neq 0$ for $R(t)>R_{\rm fog}$, i.e. for $t_n < t_T$, the
dynamics is controlled by the heat transport.

\subsection{Limit of  $\kappa = 0$}

Limit $\kappa =0$ is again specific. In case $|n -n_{P,max}
|/n_{P,max} \ll 1$, i.e. in the vicinity of the  point
$(n_{P,max},\tilde{s}_{P,max})$, corresponding solutions
(\ref{v-t}) describing dynamics of the density in the fluctuation
can be searched in the form  (\ref{delr}) with the only difference
that $\delta T$ should be replaced by $\delta\tilde{s}$,
$\lambda_{cr}$ by $\lambda_{P,max}$ and $v(T)$ by
$v(\tilde{s})$. Thus
 \begin{equation}\label{delrS} \delta n (t,r)\simeq
\frac{v(\tilde{s})}{m}\left[\pm\mbox{th} \frac{r-R_{n}
(t)}{l}+\frac{{\epsilon}}{2\lambda_{P,max}
v^3(\tilde{s})}\right]+(\delta n)_{cor},
\end{equation}
where $l=\sqrt{2c/[\lambda_{P,max} v^2(\tilde{s})]}$, the
upper sign corresponds to the growth  of the bubble and the lower
one, to the growth of the droplet,
and the solution is valid for $|\epsilon/[\lambda_{P,max}
v^3(\tilde{s})]|\ll 1$, i.e. for $\gamma_{\epsilon} =$ const $
\ll 1$ and for $R_n (0)
>R_{cr}(\tilde{s})$.
Deriving Eq.~(\ref{delrS}) we used that $\tilde{s} $ does not
depend on $t$ and $\mathbf{r}$.

Dynamics of $R_n (t)$ is determined by Eq. (\ref{dim}), where one
should replace values calculated at $(n_{cr}, T_{cr})$ to the
corresponding values at $(n_{P,max},\tilde{s}_{P,max})$.

From   Eq. (\ref{v-tS}) we obtain
\begin{equation}
\delta\tilde{s}=0 =(\delta s n_{P,max}- s_{P,max}\delta
n)/n_{P,max}^2
\end{equation}
with $\delta n$ given by Eq. (\ref{delrS}). Knowing $\delta s$ and
$\delta n$ we find $\delta T$ from Eq. (\ref{T-s}). Thus not only
density, $n$,  but also the entropy density, $s$, and the temperature, $T$,
change in the surface layer and get different values inside and
outside the bubble/droplet. Contrary, the value $\tilde{s}$
remains constant.

\section{Dynamical effects of the expansion} Let
$[t_{exp}(n)]_{\tilde{s}}$ is the typical time for the expansion
of the system (at approximately constant entropy) prepared in
heavy ion collision and the system trajectory passes through the
ITS  region.  In case, if $t_{aer}> [t_{exp}(n)]_{ \tilde{s}}$ the
aerosol has no time to develop in the corresponding density
interval. Then the system (being covered by fine ripples)
continues a quasi-uniform expansion traversing the ITS region.
Further it may cross the AS region (note that it also can be that
the expanding system freezes out before the trajectory crossed the
AS line). As is seen from Fig. 2,
the turning line can be further reached, where $P(n_{t})=0$. For
$\tilde{s}>\tilde{s}_{t1}$ the turning points are lying inside the
AS region. For $\tilde{s}_{t2} <\tilde{s}<\tilde{s}_{t1}$ the
turning points are outside the AS region, but within the ITS
region. In the vicinity of the turning point, the expansion time
$t_{exp}(n_{t})$ can be significantly increased yielding
$t_{aer}<t_{exp}(n_{t})$. Note that, as follows from the analysis
with the realistic equations of state the turning points can be
reached only for low values of the entropy $\tilde{s}$. Such low
values are reachable in case of low energy heavy ion collisions,
where there arises nGL transition. In case of violent heavy-ion
collisions at conditions of the first-order hQGP phase transition
values of $\tilde{s}$ are much larger and $P$ is always positive.

\section{Concluding remarks}
In this work  we found various
solutions describing dynamics of the density, temperature, and
entropy density in the spinodal and metastable regions
at first-order phase transitions.
We tried to answer the question 
how one should treat the fact that critical temperatures at the
isothermal and adiabatic lines are different. In particular, we
demonstrated that correct region of instability is not
appropriately found within standard ideal hydrodynamical calculations (where instability manifests itself at
the crossing of the adiabatic line). This
might be  an important finding since ideal hydrodynamics is
often used in
actual simulations of heavy ion collisions. Since  in reality
$\kappa \neq 0$, spinodal instabilities  take place, when the system
phase trajectory crosses the
isothermal spinodal line in the course of the heavy ion collision. The value of the critical
temperature, $T_{cr}$, at the pressure isotherm [$P(n)$ for fixed
$T$] is significantly larger than the value of the maximum
$T_{P,max}$ at the pressure adiabat [$P(n)$ for fixed
$\tilde{s}$]. This means that according to  our findings the
prospects of  observing signatures of spinodal instabilities in
experiments with heavy ions become more promising.

\vspace*{5mm} {\bf Acknowledgements} \vspace*{5mm}

We are grateful to  Yu.B. Ivanov, B. Friman and E.E. Kolomeitsev
for fruitful discussions.
We are
grateful  to L. Grigorenko for the reading of the manuscript
and making numerous useful comments.
V.~S. acknowledges the financial support by the Frankfurt Institute for Advanced Studies (FIAS).
This work was supported in part by the
DFG grants WA 431/8-1 and 436 RUS 113/558/0-3.

{\bf {Appendix A.} Van der Waals equation of state}

All thermodynamical
quantities can be found from expression for the free energy
\begin{equation}\label{F} F[n,T] =Nf_0 \left\{ T\left[ \ln
(\lambda^{3}n)-1\right]- T\ln (1-bn)-an\right\},
\end{equation}
where  $\lambda =\sqrt{2\pi/(mT)}$ is the thermal wave-length, $N$
is the baryon number. Parameter $a$ governs the strength of a mean
field attraction and $b$ controls a short-range repulsion, $f_0$
is an extra constant, which can be fitted to obtain more realistic
values of thermodynamic quantities for the specific phase
transition (e.g., for nGL or hQGP). From Eq. (\ref{F}) we find the
pressure in $(n,T)$ variables:
\begin{equation} \label{P} P[n,T]=f_0\left[ {nT}/{(1-bn)}-n^2 a \right],
\end{equation}
and the entropy per baryon
\begin{equation} \label{S} \tilde{s}[n,T]=f_0\left[ -\ln (\lambda^3/b)+\ln
(1/(bn)
-1)+5/2\right]+C_1.
\end{equation}
Since entropy  in classical systems is defined up to a constant,
an arbitrary constant $C_1$ is added.

From Eqs. (\ref{P}) and (\ref{S}) we also find the energy
\begin{equation} E[n,\tilde{s}]=Nf_0 (3B x^{2/3}-an),
\end{equation}
the pressure
\begin{equation}\label{Ps} P[n,\tilde{s}]=f_0 (2B x^{5/3}- an^2 ),
\end{equation}
and the temperature
\begin{equation}
T[n,\tilde{s}]=2Bx^{2/3},
\end{equation}
expressed in terms of $(n,\tilde{s})$ variables.  Here
$B=(\pi/m)\mbox{exp}[2\tilde{s}/(3f_0)-5/3 +C_1]$, $x=n/(1-bn)$.
In numerical calculations, which results are illustrated by Figs.
1,2, being performed just with a demonstration purpose we put $f_0
=1$ and $C_1 =0$, as for the case of the standard Van der Waals
equation of state.

{\bf {Appendix B.} Spinodal instability in  relativistic
hydrodynamics}

In this Appendix we generalize  results obtained in Sect.
\ref{Instabilities} to the relativistic case. We use the Landau
approach~\cite{Landau}. It is well-known that equations of
relativistic first-order hydrodynamics are parabolic and formally
suffer of the causality problem. However
 the causality problem may arise only, when
one tries to describe the flow with a fast velocity and for short
wavelength  phenomena.  To avoid this problem  following
Ref.~\cite{Minami:2009hn} we assume that the velocity of the
collective motion is much slower than the mean particle velocity
in the system (the latter can be relativistic) and we are
interested in description of a long wavelength  phenomena.

 Let the equilibrium state be the
fluid rest frame. Then $\delta u^\mu =(0,\delta \mathbf{u})$. As
in non-relativistic case we linearize the hydrodynamical equations
with respect to the variations $\delta n$, $\delta s$, $\delta T$
and  $\delta \mathbf{u}$. With the help of the standard
thermodynamic relations the Landau equations are reduced to
~\cite{Minami:2009hn}:
\begin{eqnarray}
\label{rel_hydro1} && \partial_t \delta n + n (\mathbf{\nabla}
\delta\mathbf{u}) - \kappa \frac{nT}{w^2} \nabla^2\left( \delta P
- c n \nabla^2 \delta n - \frac{w}{T} \delta T \right)=0, \\ && w
\partial_t (\mathbf{\nabla} \delta\mathbf{u}) -\nabla^2\left[   ( \zeta
+ \tilde{d} \eta )  (\mathbf{\nabla} \delta\mathbf{u})  - \delta P
+ c n \nabla^2\delta n\right] = 0, \\ && \partial_t \delta s -
\frac{s}{n} \partial_t \delta n + \frac{\kappa}{w}\nabla^2\left(
\delta P - c n \nabla^2\delta n - \frac{w}{T} \delta T   \right)
=0, \label{rel_hydro3}
\end{eqnarray}
where $w=\epsilon + P$ and $s$ are the enthalpy density and the
entropy density, respectively. Here we additionally included
surface tension effects.

The pressure and the entropy density variations are expressed
through the density and the temperature variations (see e.g.
Eq.~(\ref{varP}) for the pressure variation).
Substituting Eqs.~(\ref{delns}) in
Eqs.~(\ref{rel_hydro1}-\ref{rel_hydro3}) we find the dispersion
law. In the long wavelength limit (up to $p^2$-terms) we obtain
\begin{eqnarray}
&&\gamma_{1,2}^{\rm rel} = \pm i c_{\tilde{s}} p + \left[
\frac{\kappa}{c_V}\left(\frac{u_T^2}{u_{\tilde{s}}^2}-1 \right)  -
(\zeta+\tilde{d}\eta) \right]\frac{p^2}{2}-
\frac{c_{\tilde{s}}^2T}{w} (2\alpha-1)\kappa\frac{p^2}{2}, \\
&&\gamma_3^{\rm rel} = - \frac{\kappa u_T^2}{u_{\tilde{s}}^2 c_V}
p^2 \label{rel_disp}.
\end{eqnarray}
Here $c_{\tilde{s}}^2=(\partial P/\partial
\epsilon)_{\tilde{s}}=(n/w)(\partial P/\partial
n)_{\tilde{s}}=(mn/w) u_{\tilde{s}}^2$, $\alpha=c_V/c_P = u_T^2
u_{\tilde{s}}^{-2}$. Similar to the non-relativistic case, the
solutions $\gamma_{1,2}^{\rm rel}$ describe  sound excitations in
the long wavelength limit, while $\gamma_3^{\rm rel}$ corresponds
to the thermal mode. The last term in expression for
$\gamma_{1,2}^{\rm rel}$ is a small correction $\propto O(T/m)$
for $T\ll m$ of our interest. Dropping it we arrive at Eqs.
(\ref{solG1}), (\ref{solG}), which we have derived in  the
non-relativistic limit in the paper body. Thus we come to
precisely the same conclusions about instability region, as in the
non-relativistic case.

\end{document}